\documentclass[12pt]{elsarticle}

\usepackage[margin=2.54cm]{geometry}
\usepackage{graphicx}
\usepackage{svg}
\usepackage{hyperref}
\usepackage{url}
\usepackage{placeins}
\usepackage{multicol}
\usepackage{textcomp}
\usepackage{stfloats}
\usepackage{enumitem}
\usepackage{xcolor}

\usepackage{amsmath,amssymb,amsfonts}

\usepackage[linesnumbered, ruled, vlined]{algorithm2e}
\SetKwInOut{Variables}{Variables}

\SetCommentSty{mycommfont}
\makeatletter
\newcommand*{\compress}{\@minipagetrue}
\makeatother

\newcommand{\techName}{LEAPME}
\newcommand{\change}[1]{\textcolor{black}{#1}}
\newcommand{\paperTitle}{\techName{}: Learning-based Property Matching with Embeddings}

\begin{document}

\begin{frontmatter}

\title{\paperTitle}

\author[label1]{Daniel Ayala\corref{cor1}}
\ead{dayala1@us.es}
\cortext[cor1]{Corresponding author}

\author[label1]{Inma Hernández}
\ead{inmahernandez@us.es}

\author[label1]{David Ruiz}
\ead{druiz@us.es}

\author[label2]{Erhard Rahm}
\ead{rahm@informatik.uni-leipzig.de}

\address[label1]{Universidad de Sevilla\\
ETSII, Avda. Reina Mercedes, s/n. Sevilla, Spain}
\address[label2]{Leipzig University\\
Institut für Informatik. Leipzig 04109, Germany}

\begin{abstract}
Data integration tasks such as the creation and extension of knowledge graphs involve the fusion of heterogeneous entities from many sources. Matching and fusion of such entities require to also match and combine their properties (attributes). However, previous schema matching approaches mostly focus on two sources only and often rely \change{on simple similarity measurements}. They thus face problems in challenging use cases such as the integration of heterogeneous product entities from many sources.

We therefore present a new machine learning-based property matching approach called \techName{} (LEArning-based Property Matching with Embeddings) that utilizes numerous features of both property names and instance values. The approach heavily makes use of word embeddings to better utilize the domain-specific semantics of both property names and instance values. \change{The use of supervised machine learning helps exploit the predictive power of word embeddings}.

Our comparative evaluation  against \change{five} baselines for several multi-source datasets with real-world data shows the high effectiveness of \techName{}. We also show that our approach is even effective when training data from another domain (transfer learning) is used.
\end{abstract}


\end{frontmatter}

\section{Introduction}
\label{sec:introduction}

Data integration tasks such as the creation and refinement of knowledge graphs have to increasingly deal with the matching and fusion of data from many sources, e.g., different web sites, already created knowledge bases and repositories.  Such knowledge graphs (KG) physically integrate numerous entities with their properties (attributes) and relationships as well as associated metadata about entity types and relationship types in a graph-like structure \cite{rahm2016case}. Many companies (including Google, Facebook, and Amazon) are increasingly relying on the integrated and curated information in knowledge graphs and there is also an increasing amount of research on KG creation ~\cite{dong2014knowledge,szekely2015building,shi2017semantic,xu2019cross,ayala2019aynec,borrego2019generating,obraczka2019,saeedi2020} and KG exploitation, e.g. for question answering ~\cite{zheng2018question,huang2019knowledge}.

Integrating new data sources and their entities into a KG is challenging due to the typically large number of different kinds of entities and relationships, the high degree of heterogeneity in their representations and the often low data quality with frequently incomplete, wrong or contradicting information. Subproblems to deal with include the categorization, matching, clustering  and fusion of entities. These steps in turn also require to match the properties  of entities, e.g., to focus entity matching on comparable properties or to fuse the values of equivalent properties. 

\begin{figure*}[!ht]
    \centering
	\includegraphics[width=1\linewidth]{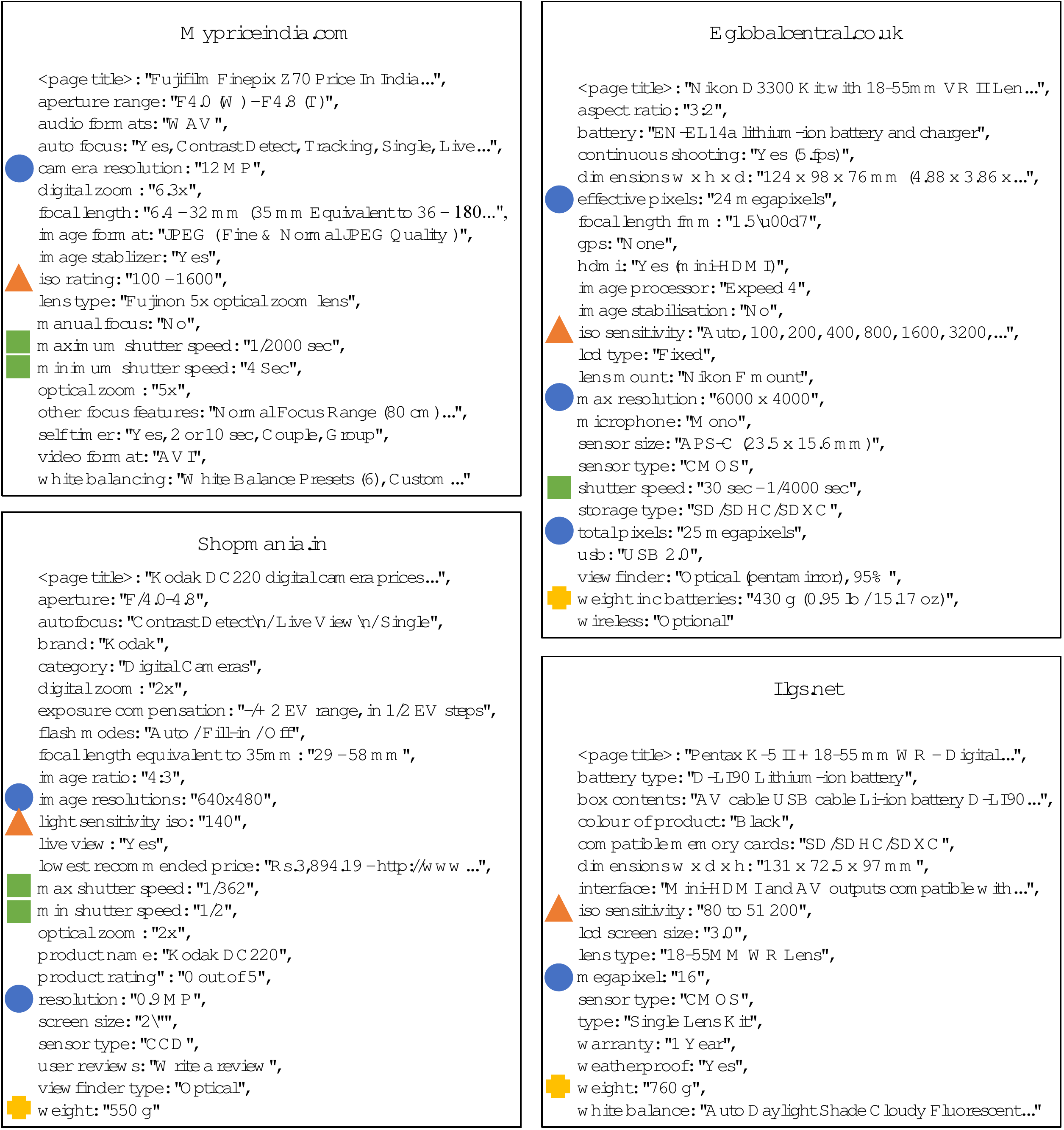}
	\caption{Camera properties from different sources. Two properties from different sources being annotated with the same shape denotes a match.}
	\label{fig:example}
\end{figure*}

Matching properties is far from  trivial, especially with many sources. As an example, Fig.~\ref{fig:example} shows camera entities (from a real dataset used in our evaluation) from four sources that may be integrated into a product KG together with some property matches indicated by symbols of the same shape. The example shows that there are numerous similar but differently named properties with diverse instance values. Matching properties often have completely different names, e.g., for properties ``camera resolution",  ``effective pixels"  and ``megapixel".
A property in one source, e.g. ``shutter speed", may also have several matches in another source, e.g., ``min shutter speed" and ``max shutter speed". 
The instance values also show a high degree of heterogeneity due to the use of synonyms, abbreviations, different technical units, or numeric values, making it difficult to find matches with standard techniques that rely on string similarity metrics \change{applied to either property names or instances. Even if more sophisticated techniques are used (e.g. word embeddings), the computation of similarities is usually unsupervised, making it hard to set thresholds that consistently achieve a high similarity for related properties and a low one for unrelated ones}. The number of  properties per entity may also differ to a large degree between sources and even within a source, which may affect some techniques.  

To help solve the property matching problem in the case of such scenarios, we present a new approach called \techName{} (LEArning-based Property Matching with Embeddings). It uses \change{supervised} machine learning and makes use of the typically good availability of instance in a KG. \techName{} applies a dense neural network and a large set of features to classify a pair of properties from different sources as related or not.
The proposed features make heavy use of word embeddings computed from both the property names and their instance values. Word embeddings are numeric vectors associated to single words, created so that they preserve their semantics. The use of embeddings gives the classifier information about the semantic proximity between two properties even when their string similarity is low. For example, we expect different words related to camera resolution such as ``MP", ``resolution" or ``megapixels" to have similar embedding vectors. The use of property values provides additional information that is not tied to the name of a property, and makes the proposal applicable to scenarios in which the properties do not have meaningful names, e.g., identifiers that are automatically generated by information extraction approaches~\cite{roldan2019extracting}. \change{The use of machine learning helps use these features in a smart way, learning what features are more important and how they must be combined, which is of great relevance when it comes to word embeddings, since they can have a high number of components that would make setting manual weights and similarity thresholds very difficult.}

Specifically, we make the following contributions:
\begin{itemize}
\item We propose \techName{}, a new learning-based approach for property matching that is applicable for data integration in scenarios with many sources \change{that result in a high degree of heterogeneity}, e.g., as needed for KG creation and refinement. We propose the use of numerous features derived from both the property names and property values and the heavy use of word embeddings for high match quality. \change{These features are exploited by a supervised classifier to avoid setting manual weights and similarity thresholds for such features.}
\item We comprehensively evaluate \techName{} on four real-world datasets with entities from several e-commerce contexts. \change{The multi-source setting makes it reasonable to use some sources as training data in order to match the rest}. We also provide a comparison with \change{five} baselines and show that \techName{} clearly outperforms previous approaches even with little training data.
\item  We  show that \techName{} can also achieve better results than the baselines when trained with data obtained from entities of a different type or domain. The reduced need for domain-specific training data increases the applicability and impact of the new approach, \change{which shows that the application of supervised machine learning and the development of labelled property matching datasets including varied domains is crucial, since they allow the creation of context-independent universal classifiers.}
\end{itemize}
The next section describes related work on schema matching and the previous use of machine learning for this task. \change{Section~\ref{sec:problem-definition} formally describes the problem of property matching}.  Section~\ref{sec:our-approach} describes \techName{} in detail. Section~\ref{sec:evaluations} contains the evaluation including the comparison with several baselines and the use of transfer learning. Section~\ref{sec:conclusions} summarizes our contributions and discusses potential future work.
\section{Related work}
\label{sec:related-work}

In the last decades, a huge amount of research has been devoted to schema and ontology matching to automatically determine corresponding schema attributes (properties) and ontology concepts. As described in several survey articles and books~\cite{rahm2001survey,shvaiko2007ontology,bellahsene2011schema,bernstein2011generic,otero2015ontology}, most of the proposed approaches focus on pairwise matching between two schemas or ontologies and utilize a combination of several similarity values to determine likely matches. The most common approach is to determine the linguistic similarity of properties either based on string similarity metrics, synonym information from background knowledge resources such as dictionaries (e.g., WordNet~\cite{lin2008survey}), \change{or, more recently, pre-trained word embeddings~\cite{fernandez2018seeping,kolyvakis2018deepalignment}. Background knowledge resources can even include a corpus of formerly matched schemas as support for a new match~\cite{madhavan2005corpus,rahm2011towards}}. Some approaches additionally utilize the structural similarity of elements (e.g., based on the similarity of neighbors in an ontology) and the similarity of associated instance data\change{~\cite{qin2007discovering,duan2012instance}}.  

\change{Taking these considerations into account, we have used four of the existing pairwise tools as unsupervised baselines in our comparative evaluation according to their reported performance or similarity to our proposal. Two of them, Agreementmaker Light (AML)~\cite{faria2019aml} and FCA-Map~\cite{chang2019fcamap} because of their good results in the OAEI (ontology alignment evaluation initiative). The proposal by Duan et al.~\cite{duan2012instance} because of its use of property instances, and SemProp~\cite{fernandez2018seeping} because of its use of word embeddings. }They do not use \change{supervised} machine learning to learn optimized similarity thresholds but require the user to fine-tune parameters manually or with the help of some technique~\cite{rivero2020selecting}. In particular, AML compares property names by doing a full-name match and computing word similarity, string similarity, and WordNet similarity. If any of the matchers returns a similarity above a user-given threshold (0.6 by default), the pair is considered a match. FCA-Map applies lexical matching to properties based on exact token co-occurrence. \change{The technique by Duan et al. uses local sensitive hashing to estimate the similarity between two groups of instances. SemProp uses word embeddings to identify when two concatenated property names have semantic coherence.}

The use of supervised machine learning is being increasingly applied for a simplified configuration of schema and ontology matching, \change{since it can be considered a way to aggregate several similarity metrics or matchers, removing the need to set manual thresholds or use vector distance metrics such as the cosine similarity, which give the same weight to all features}~\cite{spohr2011machine,nezhadi2011ontology,shenoy2012nn,ichise2008machine,eckert2009improving,djeddi2013ontology,shenoy2012nn,curino2007x,marie2008boosting}. The training data consists of the similarity of matching and non-matching pairs of schema/ontology elements together with multiple similarity values, e.g., according to different linguistic and structural similarities. Surprisingly, instance similarities have not been utilized so far in these approaches. As a representative baseline we consider the approach of Nezhadi et al.~\cite{nezhadi2011ontology} in our evaluation. It uses 12 name string similarity metrics as well as metrics derived from background knowledge by computing the distance of two concepts in the WordNet graph, and structural metrics based on the propagation of name similarities. They also considered 5 classification alternatives to determine matches and found out that an AdaBoost aggregation of Decision Tree classifiers achieves the best results.

The main limitation of supervised machine learning techniques is the need for training data. There are two main ways to deal with this requirement: manual provision of training data or the use of transfer learning. Manually labelling selected pairs of properties or concepts is of course laborious and does not scale well. This approach thus has to be limited to relatively small amounts of training data. With transfer learning the goal is to obtain the training from another domain or use case to avoid the provision of specific training~\cite{eckert2009improving}. This is especially valuable for scenarios such as knowledge graphs, in which there are typically already integrated entities and properties from different sources so that matching information can likely be reused. In our evaluation, we will consider both approaches: the use of manually defined training matches as well as transfer learning.

Most previous work focuses on pairwise schema and ontology matching for two sources~\cite{rahm2011towards} while we have to deal with an arbitrary number of sources with different sets of properties per entity type. While multi-source property matching also builds on pairwise property matching, the degree of heterogeneity and thus the difficulty to achieve good match quality increases with more sources. In our approach, we will determine pairwise similarities between properties that can be maintained in a similarity graph of properties from several sources. Such a graph can be used as input for clustering so that all matching properties are in the same cluster that can be used as a basis to fuse these properties. Property clustering is beyond the scope of this paper but can be done with similar algorithms to those used for clustering entities based on a similarity graph, e.g.,  \cite{hassanzadeh2009framework,saeedi2018using}. \change{Other similar approaches have been proposed to refine the initial matching~\cite{megdiche2016extensible,roussille2018holontology}}.
\section{\change{Problem definition}}
\label{sec:problem-definition}
\change{
We first provide some preliminary definitions in Section~\ref{sec:preliminaries}, then we describe the problem we focus on in a formal way with well-defined input and output in Section~\ref{sec:definition}.}

\subsection{\change{Preliminaries}}
\label{sec:preliminaries}
\begin{description}

    \item[Source:] A source $S$ is a location from where information comes, e.g., a website, a relational database, or a SPARQL endpoint, among other examples. It typically conforms to some kind of ontology and may contain structured entities of several types or classes. The example in Figure~\ref{fig:example} contains camera entities from four different sources, namely e-commerce websites such as ``Mypriceindia.com" and ``Shopmania.in". 
    
    \item[Entity and class:] An entity $e$ is a representation of something that can be uniquely identified, usually corresponding to some real world object. Entities belong to a certain source and a source-specific type or class $C$, and we denote the source and class  of entity $e$ with $S(e)$ and $C(e)$, respectively.
    The rectangles in Figure~\ref{fig:example} correspond  to different entities of type "camera": ``Fujifilm Finepix Z20", ``Nikon D3300", ``Kodak DC220", and ``Pentax K-5 II". Entities consist of several properties and their values.
    
    \item[Property and instances:] A property  is an attribute to describe information about entities. The values of a property are literals known as instances. 
    Our algorithm processes a collection of property instances represented as tuples $(p,e,v)$ where $p$ is the property name, $e$ is the entity (identifier), and $v$ is the property value. An example instance is $($``camera resolution"$,$ ``Fujifilm Finepix Z20"$,$``12 MP"$)$. The components of a property instance $i$=$(p,e,v)$ are denoted by $p(i)$, $e(i)$, and $v(i)$. Each such tuple is implicitly tied to the originating class $C(e(i))$ and source $S(e(i))$.
    
    \item[Class schema:] For the sake of flexibility in applications such as E-commerce, we do not assume the existence of a predefined schema with a fixed set of properties per class. Rather, we view the schema of class C as the collection of all differently named properties for entities of class C in the respective sources. Individual entities may use any subset of these class properties.  

    \item[Property matching:] Task of determining correspondences between the properties of different class schemas from different sources.
    
\end{description}

\subsection{Definition}
\label{sec:definition}

We address property matching for properties of the same class (e.g., camera properties). We consider the case of multi-source matching so that properties may relate to entities from an arbitrary number of sources. Correspondences are not limited to equivalence relationships but also to more complex relationships between semantically related properties. A property in one source may thus have 0, 1 or several matching properties in another source, e.g., as for property ``shutter speed"  in Figure~\ref{fig:example}. 

\change{Therefore, the problem is as follows: Given a collection or property instances $I$ corresponding to properties from $m$ sources with $m>1$, we define property matching as a binary classification problem where every pair of properties $(p_i, p_j)$ from two different sources is classified as related or unrelated. Alternatively, every pair of properties can be assigned a similarity score $sim$ indicating the strength of the relatedness}. To enable the application of supervised machine learning techniques, we also assume the provision of training data consisting of pairs of properties from different sources labelled as either matching or non-matching.

\change{The output can be represented as} a similarity graph between properties of different sources. Such a graph can be used for determining clusters of matching properties, e.g., using clustering algorithms like transitive closure or more complex approaches as in \cite{hassanzadeh2009framework,saeedi2018using}. A simple transitive closure would group all same-shaped properties in Figure~\ref{fig:example} within a cluster. This would be sub-optimal if we want to ensure that only equivalent properties are  grouped together, e.g., as useful for a fusion of property values within a KG. In such a case the properties ``min shutter speed" and ``max shutter speed" should be in separate clusters. This can be achieved with clustering techniques like in  \cite{saeedi2018using} that do not permit more than one cluster member from the same source. 
An alternative approach is to post-process the match correspondences, adopting a similar approach as in \cite{arnold2014enriching} 
that can determine the semantic type of correspondences (such as equality and part-of) and only continue with equality correspondences. The analysis of such post-processing options is beyond the scope of this paper and left for future work. 
\section{Our approach}
\label{sec:our-approach}

\change{Having defined the problem of property matching, we give an overview of our proposal \techName{} (Section~\ref{sec:overview})}, and describe in detail how features are computed (Section~\ref{sec:features}). We discuss the use of embeddings in Section~\ref{sec:embeddings}. Finally, we describe aspects related to the implementation of \techName{} in Section~\ref{sec:implementation}.

\subsection{Overview}
\label{sec:overview}
\techName{} is a \change{supervised} ML-based property matching approach that focuses on the use of novel features. It computes features from property instances, property names, and property pairs to obtain large feature vectors that can be properly handled by a classifier. For example, from the instance value ``12 MP" we can compute features such as the number of digits (2), the number of white spaces (1), or the fraction of letters (0.4). \techName{} thus uses such characteristics \textit{about} instance values (and property names) in addition to their actual values.

\begin{figure}[!htp]
    \centering
	\includegraphics[width=0.5\linewidth]{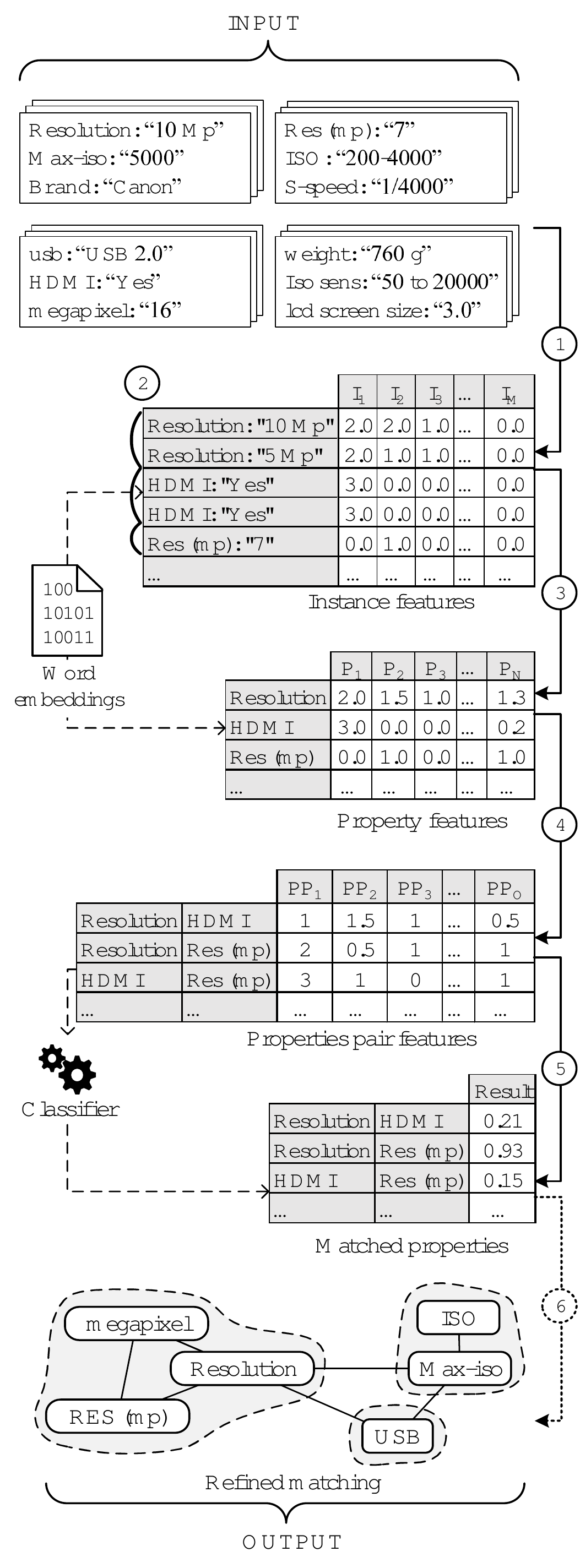}
	\caption{Workflow of \techName.}
	\label{fig:workflow}
\end{figure}

Algorithm~\ref{algo} describes the main steps  of \techName{}; the workflow is also illustrated in Figure~\ref{fig:workflow}.

\begin{enumerate}
  \item First, there is the initialization of the instance feature vector $IF$, the property feature vector $PF$, the property pair feature vector $PPF$, and the output similarity graph (collection of matches) $Sim$ (line 1 of Algorithm~\ref{algo}).  
  \item Next, the  instance features are determined by every instance  with the help of function $iFeatures$, and added to the respective property in the instance feature vector $IF$  (lines 2-3 of Algorithm~\ref{algo}, step 1 in Fig.~\ref{fig:workflow}). The features we determine will be described below - they include meta-features about the instance values as well as an embedding vector for the specific property value.
  \item  In lines 4-6 of Algorithm~\ref{algo} we compute property features with the help of function $pFeatures$  (steps 2 and 3 in Fig.~\ref{fig:workflow}). They can be derived for the property name or based on the aggregation of instance features, e.g., average values of numeric instance features.
  \item For each property pair, we compute the property pairs features using function $ppFeatures$ (lines 7-9 of Algorithm~\ref{algo}, step 4 in Fig.~\ref{fig:workflow}), which may be partially based on the aggregation of property features.
  \item 
  We use the input training data with their labeled property pairs and associated feature vectors to train a classification model using function $trainClassifier$ in line 10 ($labeled(PPF)$ denotes the already labeled  property pairs). Then, we apply the trained classifier to the unlabeled property pairs to obtain a match decision and similarity score for each pair (lines 11-12 of Algorithm~\ref{algo}, step 5 in Fig.~\ref{fig:workflow}).
\end{enumerate}  
As shown in the last step  of Fig.~\ref{fig:workflow}, the output represents a similarity graph that can be post-processed as discussed above.

\renewcommand{\in}{\text{~\bf in~}}

\begin{algorithm}[!ht]
    \label{algo}
    \DontPrintSemicolon
    \KwIn{
      \\\noindent - $I$: set of property instances from $m$ sources
      \\\noindent - labeled property pairs (training)

    }
    \KwOut{
      \\\noindent - $Sim$: set of  property pairs with  similarities (similarity graph)
    }
    \Variables{
      \\\noindent - $IF$: Map$\textless Property,FeaturesVectorSet\textgreater$ with instance features vectors, grouped by property
      \\\noindent - $PF$: Map$\textless Property,FeaturesVector\textgreater$ with property features vectors.
      \\\noindent - $PPF$: Map$\textless PropertyPair,FeaturesVector\textgreater$ with property pair features vectors.
      \\\noindent - $m$: classification model
    }
  $\text{initialize}(IF, PF, PPF, Sim)$\;
  \tcp{Steps 1-4: compute features}
  \For{$i \in I$} {
    $IF[p(i)] \gets IF[p(i)]\cup\text{iFeatures}(i))$\;
  }
  \For{$(p, V) \in IF$}{
        $PF[p] \gets \text{pFeatures}(p)$\;
  }
  \For{$p_1 \in \text{keyset of }PF$}{
    \For{$p_2$  from different source $\in \text{keyset of }PF$}{
        $PPF[(p_1, p_2)] \gets \text{ppFeatures}(p_1, p_2)$\;
    }
  }
    \tcp{Step 5: training and classification}
    $m \gets \text{trainClassifier}(labeled(PPF))$\;
    \For{$(p_1, p_2):v \in unlabeled(PPF)$} {
        $Sim.add((p_1, p_2, m.\text{classify}(v)))$\;
    }
  \caption{\techName{}}
\end{algorithm}

\subsection{Features}
\label{sec:features}
    
    Since we classify pairs of properties, the features that are ultimately fed to the classifier must be associated to a pair of properties. However, as we have mentioned, \techName{} considers features at several levels that can be later transformed into property pairs features. Next, we describe in detail each of these levels:

\begin{description}
    \item[\textbf{Instance features:}] These features are computed from each individual instance of a property (that is, a features vector is obtained for each property value) independently of the property names. They provide information about the format of property values and can thus be considered as meta-features. We expect matching properties to follow similar formats, which should be reflected in these features. For example, while in Figure~\ref{fig:workflow} properties ``Res(mp)" and "megapixel" have a different name, both have short values with numeric characters, which could be reflected in features that measure the number of such characters or token types. While on their own these features may not be enough to properly match features (since, for example, many properties follow similar numeric formats), they could help disambiguate problematic cases. Furthermore, in some contexts the name of the properties may be unknown  or only a generic identifier. For example, information extraction techniques may identify a piece of text as an instance, but not be able to infer a label with its property name~\cite{roldan2019extracting}. In these cases, no features can be computed from the property names, and only these instance features enable matching. In addition to format-oriented meta-features we also consider the actual property values in the form of word embeddings or the numeric value (see below). 

    \item[\textbf{Property features:}] These features are computed for each individual property. They include all features computed from the property name, such as the average embeddings vector of its words. Furthermore, by grouping the instance features on a per-property basis, we can aggregate them and turn them into property features. For example, we could compute the average of each instance feature for a given property to represent  the overall format followed by its instances.
    
    \item[\textbf{Property pair features:}] These features are computed  for each pair of properties to be classified. These are the final features actually fed to the classifier. Traditional string similarity metrics such as the Levenshtein or Jaro-Winkler distance would be part of these features, since they are computed from a pair or property names. Aggregated property features can also be used to determine property pair features. In this case, only two vectors are aggregated, e.g. by computing the numeric difference or average between the vectors, or by determining their concatenation.
\end{description}

\change{Note that while only property pair features are relevant to the classification of property pairs, the other features are also used but are necessarily  transformed into property pair features. For example, since a property can have hundreds of instances, there is a need to aggregate the hundreds of sets of instance features.}

\subsection{Embeddings and classification}
\label{sec:embeddings}

When matching properties, a high value of the string similarity of the property names is usually a clear indicator of a match. Low similarity, however, can be caused by the issues we mentioned in Section~\ref{sec:introduction}. As discussed in Section~\ref{sec:related-work}, the usual way to mitigate this problem is to use external knowledge bases like WordNet to determine synonyms or name-independent similarities.  These resources, however, are language-dependant and often of limited coverage. Furthermore, their use is relatively complex, and may require the use of APIs to handle the data.

As a more promising and versatile approach to overcome these limitations, we propose the use of word embeddings for both  property names and property values. They can provide  rich information about the semantics of a property that can help solve some issues such as the potentially low string similarity between synonymous properties. While embeddings are language-dependant, versions for different languages can easily be trained from any large text corpus, unlike knowledge bases such as WordNet, whose creation requires a large manual effort. Furthermore, embeddings can be trained with a context-specific corpus, and are more likely to contain certain concepts. For example, the GloVe embeddings we use contains an entry for the word ``28mm", which is a typical aperture value for cameras.

Embeddings vectors usually have hundreds of components with unknown meanings that may require nonlinear combinations to properly exploit their predictive power. For that reason, \techName{} uses a neural network for classification, which is also a popular choice in the related work \change{and is able to properly weight features even when there is a large amount of them. While word embeddings have already been used in the past as discussed in Section~\ref{sec:related-work}, they have been exploited in an unsupervised way. Unsupervised techniques that use embeddings are forced to compute the distance (usually the cosine distance) between several embedding vectors, giving the same importance to all components, which may be detrimental when the number of components is high.}

\subsection{Implementation}
\label{sec:implementation}

\begin{table}[!ht]
    \centering
	\includegraphics[width=.7\linewidth]{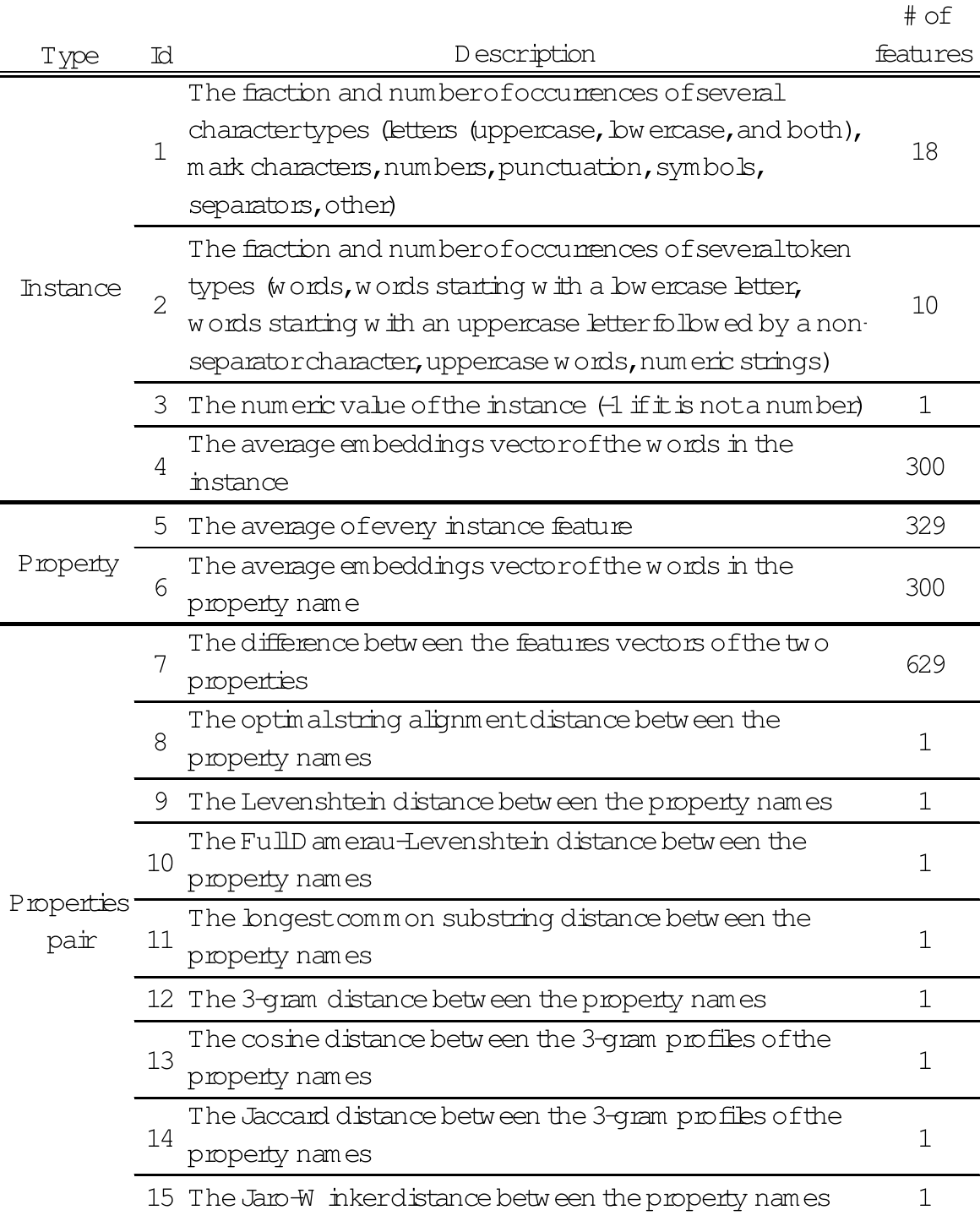}
	\caption{Features used in our implementation.}
	\label{tab:features}
\end{table}

Table~\ref{tab:features} provides an overview about the features we have implemented. Instance features are computed with TAPON \cite{ayala2019tapon,ayala2019tapon2}, which includes several format-related features to which we added the embedding ones.

\change{The rationale behind these features is the following: Features 1 and 2 contain information about the textual format of the instances, including absolute and relative frequencies of both character and token types. We expect similar properties to follow similar formats (for example, properties related to the ISO sensibility of a camera will usually contain at least 3 numeric characters). Feature 3 provides information about the specific value of purely numeric properties in order to disambiguate them according to the distribution of their values. Feature 4 provides information about the semantics of every instance. Feature 5 aggregates the instance features in order to transform them into property features by computing their average, which gives an overall idea of the format and the instance semantics of a property. Feature 6 provides information about the semantics of a property from its name. Feature 7 aggregates the property features of two properties by computing the difference of each feature in order to obtain information about the distance with regards to every feature. Features 8 to 15 use traditional string distance metrics applied to the names of the properties, since properties with similar names are usually related.}

\begin{table*}[!ht]
    \centering
	\includegraphics[width=\hsize]{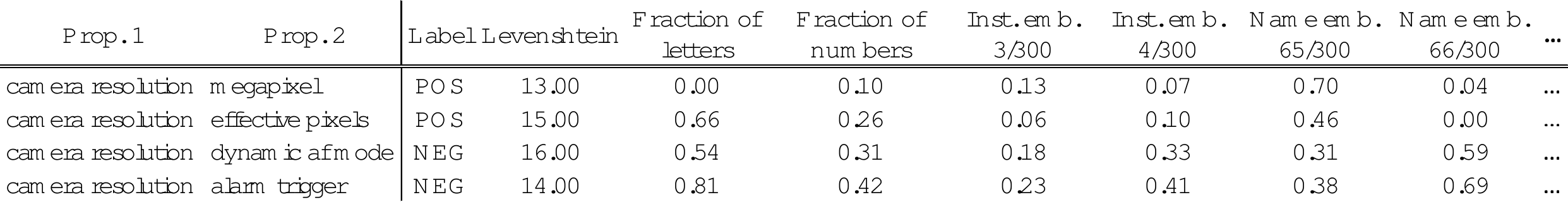}
	\caption{Sample feature values for matching and non-matching property pairs.}
	\label{tab:features-example}
\end{table*}

To compute embeddings, we use the pre-trained GloVe approach~\cite{pennington2014glove}\footnote{\url{https://nlp.stanford.edu/projects/glove/}}, specifically for the uncased Common Crawl corpus that includes 300-dimensional vectors for 1.9 million words, promising a good coverage for different domains, \change{since the corpus should contain a great variety of tokens}. Unknown words are mapped to a vector filled with zeroes. Each word in a property value or property name can thus be mapped to a point in the 300-dimensional data space so that similar words will have a small distance in it. For each property value and name we determine the average embeddings of the individual words represented by 300 values that serve as features for our classification approach. \change{We can deal with such relatively large feature vectors since the use of supervised machine learning with a neural network should be able to identify the most important ones features and give them an appropriate weight.}

As indicated in Table~\ref{tab:features}, we currently use 28 meta-features for instance values. The actual instance values are reflected in one feature for numeric values and 300 features of the embeddings vector. The averages for these 329 features over all instance values of a property serve as property features.  They are complemented with  300 features of the embeddings vector for the property name. The final feature vectors for property pairs thus include 629 features regarding the difference between the property features.
Together with 8 features about the string similarity of property names there are 637 features for property pairs in total.

Table~\ref{tab:features-example} shows an example for a small subset of our features computed for four labeled property pairs (two matching and two non-matching). Note that the Levenshtein distances are similarly high in both positive and negative match cases while other features can better  discriminate between them. For example, for feature "fraction of letters" the difference is $0$ between properties ``camera resolution" and ``megapixel" and can thus help to determine such a  match.  Some of the embedding features (corresponding to individual components of the embeddings vectors) have good discriminating potential, with a low  difference in positive  and high difference in negative match cases (e.g., instance embedding $4$ and name embedding $66$).

Finally, regarding the architecture of the neural network behind \techName{}, it consists of two fully connected hidden layers of sizes 128 and 64. We use a batch size of 32 and perform 10 epochs with learning rate $10^{-3}$, 5 with $10^{-4}$, and 5 with $10^{-5}$. We fine-tuned these hyper-parameters manually in preliminary tests, though most alterations (such as changing the size of the layers) do not significantly impact on the results. The final layer has two neurons from which the final score is obtained for the two possible outcomes (positive/negative). This allows the use of the positive output as a similarity score, which is useful for post-processing steps such as property clustering.
\section{evaluation}
\label{sec:evaluations}
We experimentally evaluate our  property matching approach \techName{} on four real-word datasets with up to 24 sources.  We analyze the impact of different amounts of training data and the effectiveness of the different kinds of features; in particular, the use of embeddings for both property values and property names. We further compare \techName{} with \change{five} baselines and study the use of transfer learning. The focus is on match quality with the standard metrics precision, recall and F-measure (F1 score).  

We first give some details about the studied feature configurations for \techName{} and the baseline approaches. Next, we describe the four datasets and the two use cases with training data from either the same or a different domain.  The results for the two use cases are discussed in subsections \ref{sec:sdresults} and \ref{sec:tlresults}, respectively. The evaluated implementations along with the detailed results and additional material are available online\footnote{\url{https://github.eii.us.es/dayala1/LEAPME}}.

\subsection{Feature configurations and baselines}

The rich set of features \change{exploited by supervised learning} is a main advantage of \techName{} and we therefore analyze the effectiveness of the different kinds of features in detail. Along one dimension, we compare the use of instance-related features only, name-related features only and the combined use of both kinds of features. Another dimension is the consideration of embedding-based features only, non-embedding features only or the combined use of both kinds of features. In total, this sums up to 9 possible feature configurations to analyze. 

The \techName{} results are compared to the results obtained by the following baselines:
\begin{itemize}
    \item The latest Github implementation of Agreement Maker Light~\cite{faria2019aml} (AML), the highest-ranked technique in the M2 variants of the ``Conferences" track of OAEI 2019, which involve the matching of only properties \footnote{\url{http://oaei.ontologymatching.org/2019/results/conference/index.html}}.
    \item The latest Github implementation of FCA-Map~\cite{chang2019fcamap}, the best-performing property matching technique in  the ``Knowledge Graph" track of OAEI 2019 \footnote{\url{http://oaei.ontologymatching.org/2019/results/knowledgegraph/index.html}}.
    \item An implementation of the machine learning proposal by Nezhadi et al.~\cite{nezhadi2011ontology}. It was selected among machine learning proposals for having the largest features catalogue, from which we removed the structural features since they were not applicable to our evaluation datasets. We use AdaBoost with decision trees as classifier, which achieved the best results in ~\cite{nezhadi2011ontology}.
    \change{
    \item An implementation of SemProp~\cite{fernandez2018seeping}, selected as a representative of existing proposals that use word embeddings. We used the proposed matchers graph by removing the use StructS, which is not applicable since we only match properties, where there are no class hierarchies involved. We tested all combinations of values $0.2, 0.4, 0.6, 0.8$ for the thresholds used by the SynM, SeMa(-), and SeMa(+) matchers. For our final experiments we used the combination that yielded the highest average F1 score across our datasets: $0.2$ for SynM, $0.2$ for SeMa(-), and $0.4$ for SeMa(+).
    \item An implementation of the proposal by Duan et al.~\cite{duan2012instance} based on local-sensitive hashing (LSH), selected as a representative of existing proposals that use property instances for matching. We tested both variants (random projections and minhash) with the proposed number of hash functions (1000 and 500 respectively) and the following band sizes: all integers from $1$ to $10$, and integers from $10$ to $50$ in steps of $5$. For our final experiments, we used the combination that yielded the highest average F1 score across our datasets: minhash with a band size of 1.
    }
    
\end{itemize}

\subsection{Datasets}
\label{sec:datasets}
For our evaluation, we use four real-word datasets with different kinds of e-commerce products (cameras, headphones, phones, and TV sets) from multiple sources. 

All datasets align the properties in each source to a reference ontology. We consider that two properties are related (matching) when they are both aligned to the same reference property. All four datasets have been extracted from the Web using information extraction techniques, and contain noise that is typical of real world scenarios, making matching more challenging.

Table~\ref{tab:datasets}  provides main statistics for the four datasets. The camera dataset  comes from the DI2KG19 challenge~\cite{DI2KG}. It 
is the largest dataset with 24 sources, more than 3200 properties and about 9200 matching property pairs. We limited the number of entities to 100 per source in order to balance their size and impact.  But, as shown in the lower part of Table~\ref{tab:datasets},  the number of different properties and the number of property entities differs substantially between different sources, with almost 700 properties for one of the sources (EBay.com). 
The other datasets contain headphones, phones and TV product entities and
correspond to the WDC Gold Standard for Product Matching and Product Feature Extraction~\cite{GSProductExtraction}. These are much smaller than the camera dataset and there are different numbers of entities per source leading to a less balanced setting than that of the camera dataset. In our analysis of the results, we will refer to the three smaller and imbalanced datasets as low-quality datasets as opposed to the high-quality camera dataset.

\begin{table}
    \centering
	\includegraphics[width=.8\linewidth]{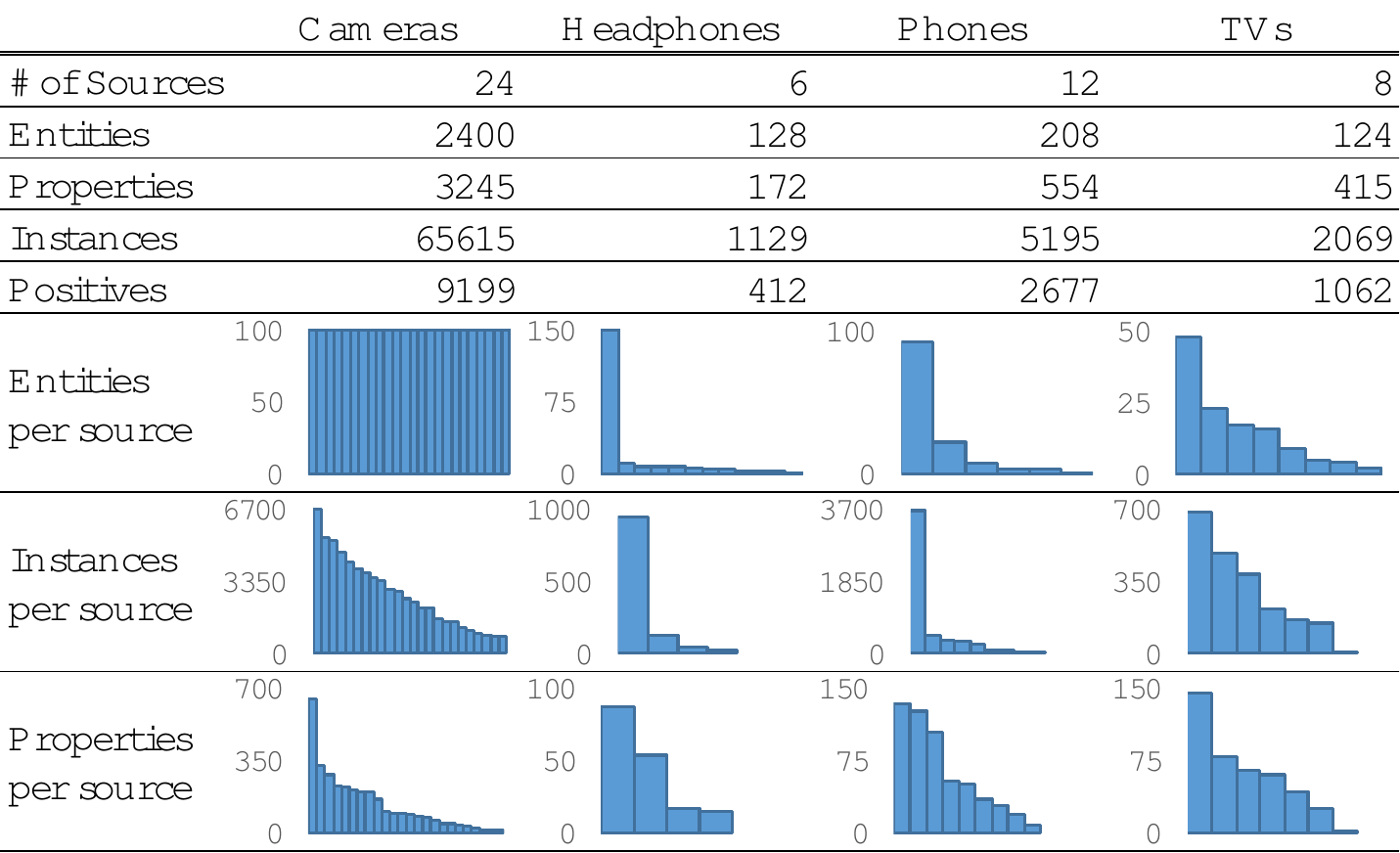}
	\caption{Datasets metadata. Each vertical bar in a plot represents a source within a dataset.}
	\label{tab:datasets}
\end{table}

\subsection{Use cases and training data}
\label{sec:training}
For training, we differentiate two use cases in which the training data either refers to properties from the same domain (entity type) or to properties from a different domain. For the first use case, which we call Single Domain (SD), we take a fraction of the sources of a dataset (at random) for training. We use the examples that involve two sources of data in the training set to train the classifier, and test it with the rest. We performed experiments using different training fractions: 0.2, 0.4, 0.6, and 0.8. For each of these fractions and for each dataset, we ran \techName{} 25 times, using different random combinations of training sources.

For the second use case, called Transfer Learning (TL), we train the classifier for one dataset (entity type) with training data from the other datasets.

For this purpose, we tested all possible combinations of using 1, 2, or 3 datasets for training. 

\begin{figure}[htp]
    \centering
	\includegraphics[width=0.5\linewidth]{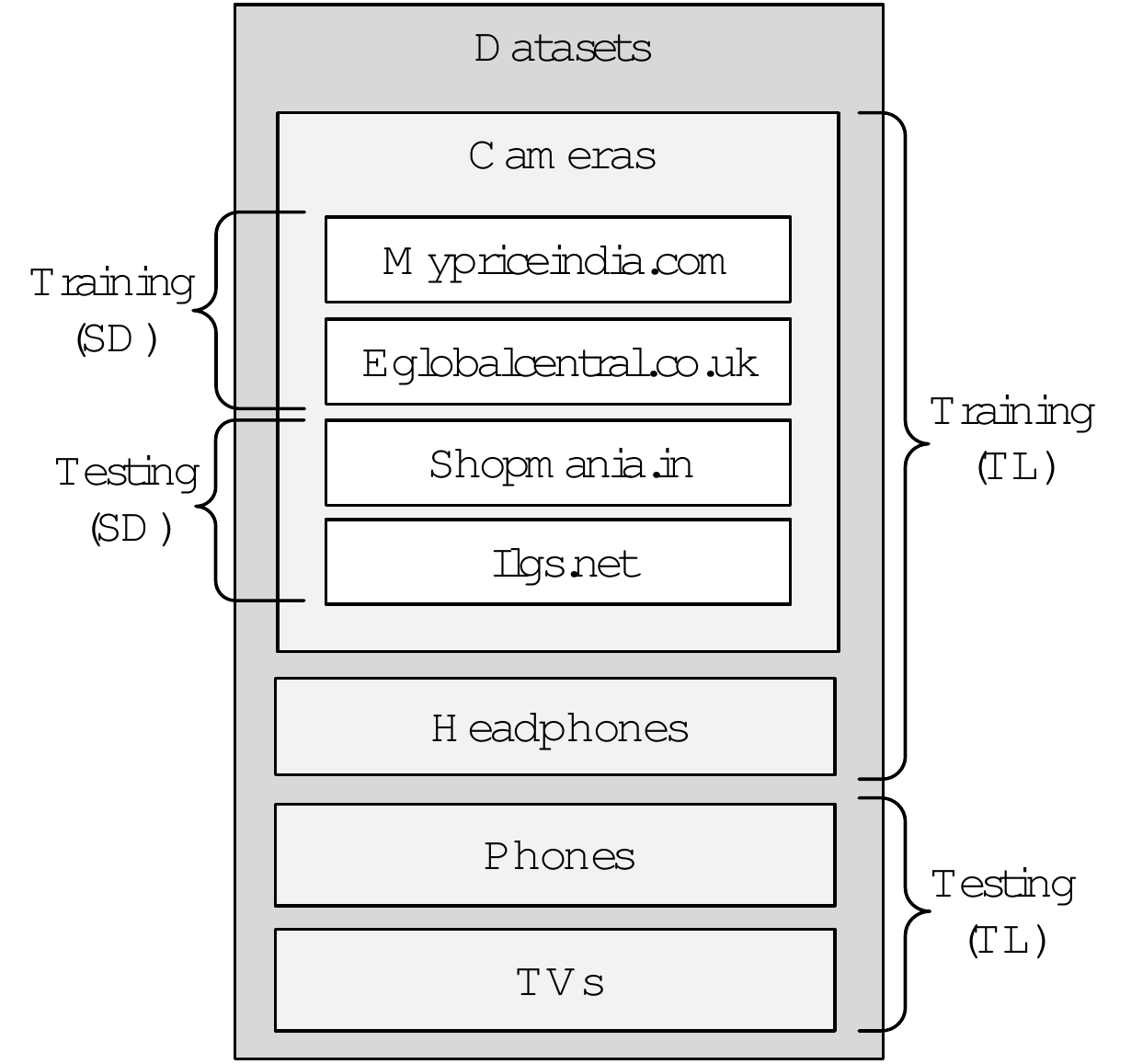}
	\caption{Use cases testing example.}
	\label{fig:use-cases}
\end{figure}

Figure \ref{fig:use-cases} shows an example of how data is divided into training and testing for both use cases. The example illustrates SD training for the camera dataset where each labeled property pair uses properties from two of the 24 sources. For the TL use case, there are many possible configurations. The shown example refers to training pairs from two datasets to be used for evaluating property matching for the rest.

For all datasets and use cases, the training data consists of two negative (non-matching) pairs of properties for every positive (matching) pair, and the negative pairs are randomly selected.

\subsection{Single-domain results}
\label{sec:sdresults}
Next, we compare the results obtained by different configurations of \techName{}, as well as those obtained by the \change{five} baselines.

\newcommand{\figwidth}{1\linewidth}
\newcommand{\figwidthEC}{0.75\linewidth}
\begin{figure*}[htp]
    \centering
	\includegraphics[width=1.0\linewidth]{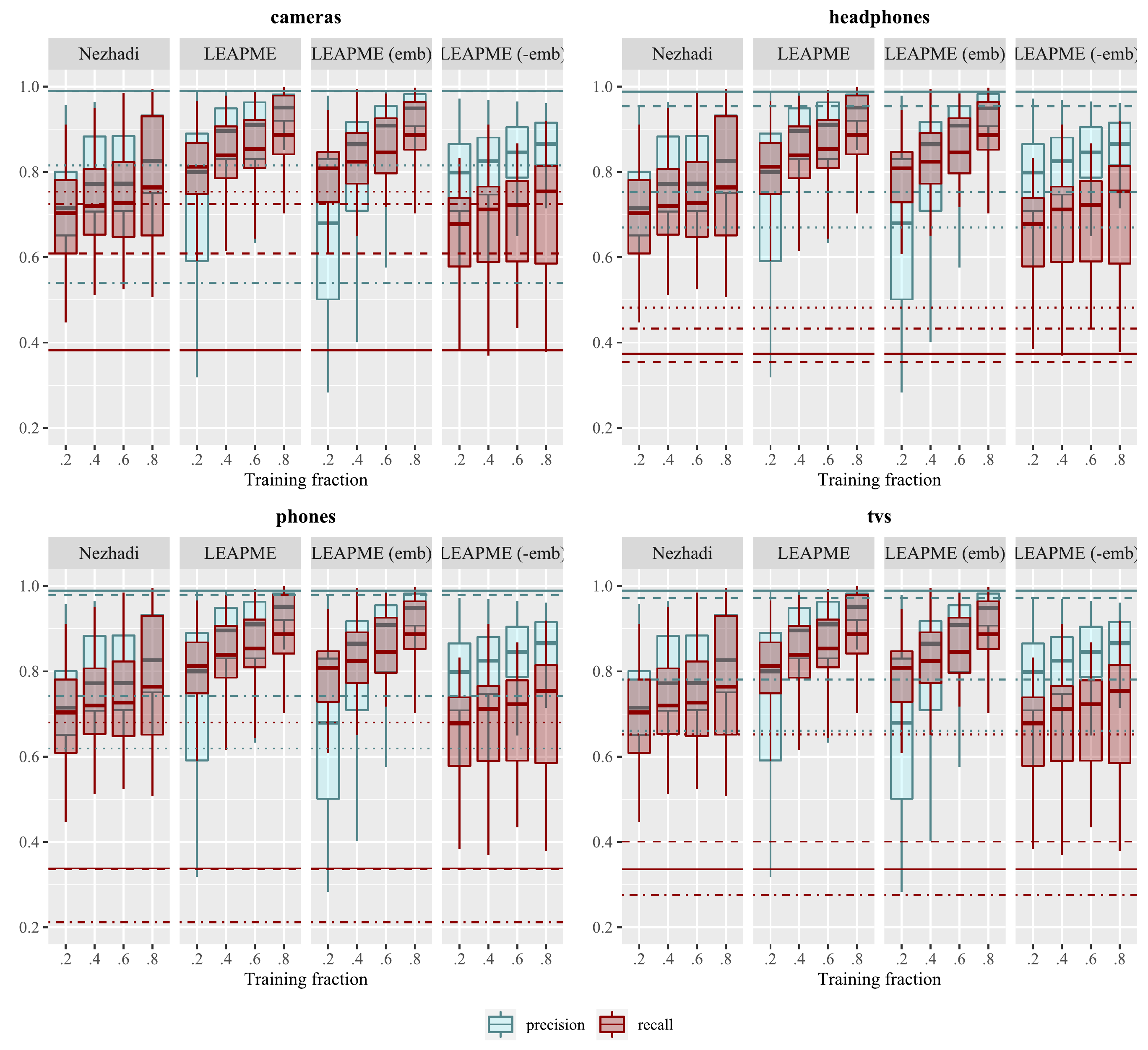}
	\caption{SD use case results. Matching with both property names and values. \techName{}(emb) = \techName{} with embedding features only. \techName{}(-emb) = \techName{} without embedding features. Dashed line = AML. Solid line = FCA-Map. \change{Dotted line = SemProp. DotDash line = LSH.}}
	\label{fig:datasets-uc1-both}
\end{figure*}

\begin{table*}
    \centering
	\includegraphics[width=1.0\linewidth]{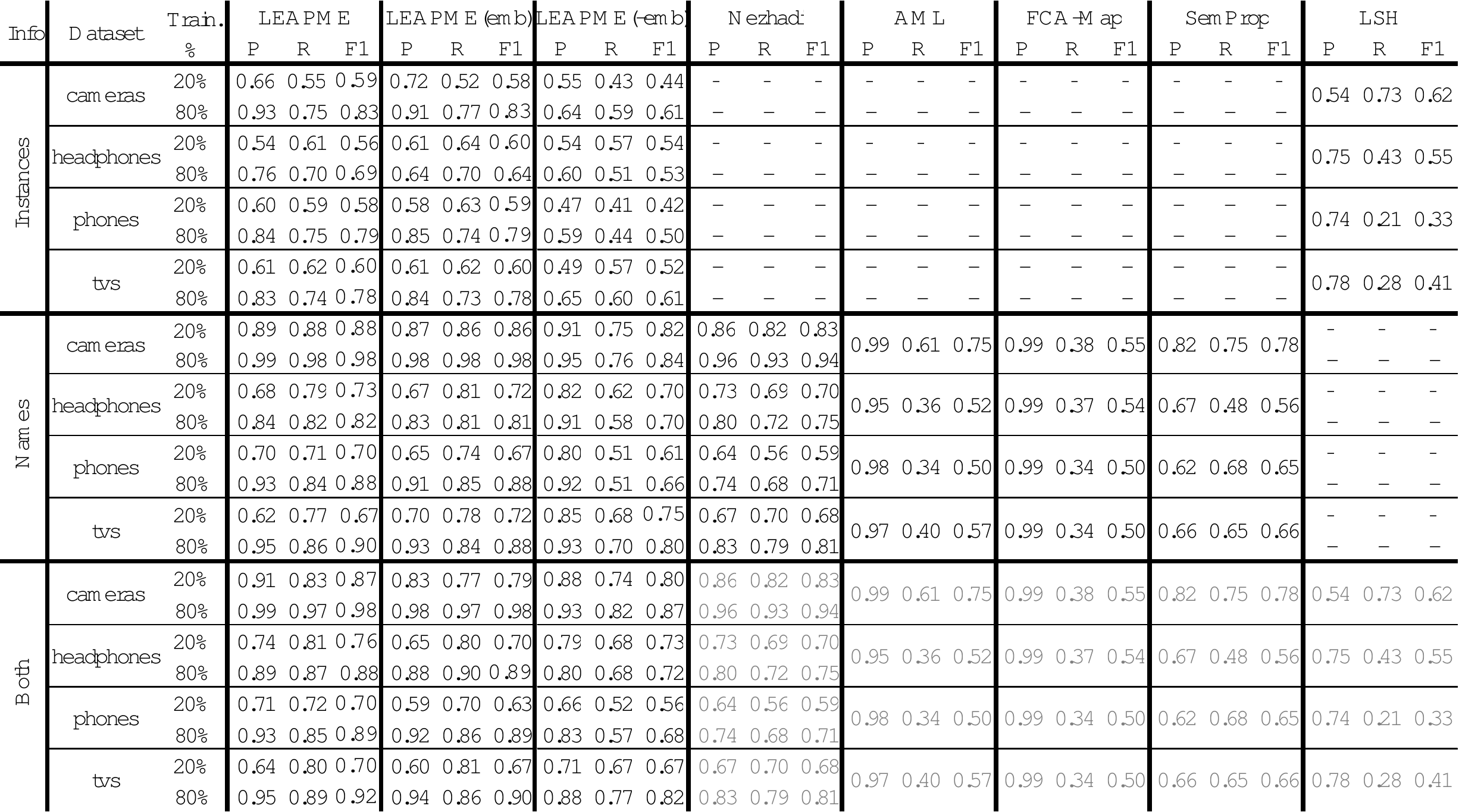}
	\caption{SD use case summary with F1 scores. \techName{}(emb) = \techName{} with embedding features only. \techName{}(-emb) = \techName{} without embedding features.}
	\label{tab:results}
\end{table*}
We first evaluate the results in the single-domain case and compare  \techName{} with the five baselines. 
Figure~\ref{fig:datasets-uc1-both} shows the precision (in blue) and recall (in red) results for the four datasets for different amounts of training data (0.2 to 0.8) for the supervised approaches \techName{} and Nezhadi. The results for the \change{unsupervised} approaches are shown as horizontal lines since they do not depend on training data. The results for \techName{} are obtained by using both features for property names and instances but we differentiate the three cases of the use of embedding features only, the use of non-embedding features only and the use of all features. The value ranges for the \change{supervised} approaches reflect the different training configurations involving a random selection of the sources in a dataset.  The variations are generally higher for the smaller and unbalanced datasets like headphones that provide fewer training data. We make the following observations:
\begin{itemize}
    \item Unsupervised techniques can achieve a high precision but struggle to reach a similar recall.
    \item \techName{} achieves better overall results than all the baselines, with a dramatic increase of recall when compared to AML and FCA-Map \change{and both recall and precision improvements when compared to the rest}. 
    \item The use of embedding features always improves recall compared to the sole use of non-embedding features, but they need enough training data to reach or surpass their precision. As expected, using all features achieves the best results for \techName{}.
    \change{\item When only using embeddings, \techName{} achieves, with $20\%$ of training data, significantly better results than SemProp, except in the cameras dataset, where results are similar. When using more training data, \techName{} achieves much better results, showing that the use of embeddings greatly benefits from supervised learning.}
 \end{itemize} 

For a more detailed analysis, we summarize the average results, including F1 scores, in Table~\ref{tab:results} for both 20\% and 80\% training data. The table also provides results for the sole use of instance features and the sole use of name features, again differentiated by the use of embedding features only, non-embedding features only or both. 
The best F1 results of each row have been marked in bold. We make the following additional observations. 

\begin{itemize}
    \item For all datasets, \techName{} achieves a better F1 score than all baseline approaches even when using only 20\% training data.  For 80\% training data, it achieves excellent F1 scores from 88\% (for headphones) to 98\% (for cameras). In this case, the baselines are outperformed especially for the low-quality and more challenging datasets (headphones, phones, and TVs). The \change{unsupervised} baselines were outperformed by up to 42 F1 percentage points  (50 vs 92\% for the TV dataset) and the \change{supervised} baseline of Nezhadi by up to 18 percentage points (71 vs 89\% for the phones dataset). 
    
    \item When only using property names \techName{} without embedding features already outperforms the baselines. The embedding features for property names are the most effective features in \techName{}. Their use alone is more more effective than the use of non-embedding features relying on string similarities. 
   
    \item Only using instance features  achieves weaker results for \techName{} than using name features especially with little training. Again, using embedding features is more effective than using the non-embedding ones that focus on format-oriented meta-features. Still, the combination of both instance and name features helps to achieve a slight improvement over the sole use of name features in most cases. 
    \end{itemize}
 
 \change{Table~\ref{tab:results} includes a number of further interesting results such as that SemProp  outperforms the other unsupervised baseline approaches due to its use of embeddings. For the TV dataset, the use of non-embedding features proved to be slightly more effective than the use of embeddings for 20\% of data for training showing that the  effectiveness of embeddings can depend on the availability of a sufficient amount of training.}

\subsection{Results for transfer learning}
\label{sec:tlresults}

\begin{figure}[!ht]
    \centering
	\includegraphics[width=.72\linewidth]{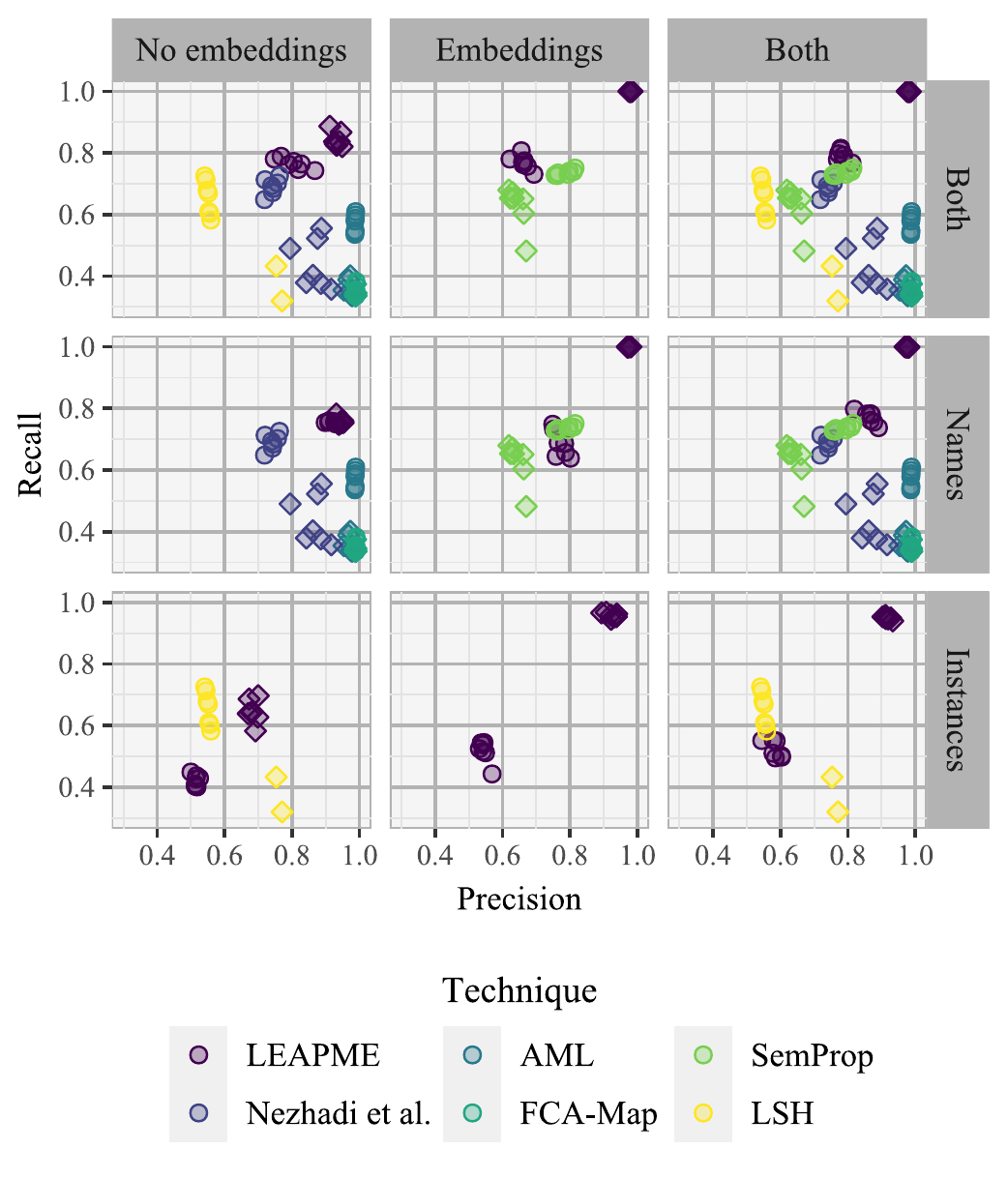}
	\caption{TL use case results. Rhombuses = high quality datasets in training split. Circles = high quality datasets in testing split.}
	\label{fig:datasets-uc2}
\end{figure}

Figure~\ref{fig:datasets-uc2} shows the precision and recall results of our experiments in the TL use case. Each point corresponds to a different combination of three or less datasets used for training, the rest being used for validation. Note that  baseline results are the same when using both name and instances or only of them, since even when both are available, they only use one. The figure also distinguishes between the use of high-quality and low-quality datasets for training. 

We observe that, similarly as in the SD use case, the unsupervised baselines can achieve high precision but suffer from a lower recall. \techName{} again achieves the best results for embedding features especially for names or both names and instances.  The best results are generally achieved when the training includes data from the high quality dataset (camera dataset) that can also provide more training samples than the low-quality datasets. \techName{} achieves near-perfect precision and recall when using it for training, demonstrating that it can make excellent use of transfer learning for good training data from a different domain. By contrast, the \change{supervised} technique by Nezhadi et al. only achieves better precision for training from a high-quality dataset but worse recall making it less suitable for transfer learning. \change{SemProp can achieve results similar to \techName{}, but only when low quality training data is used}. This indicates that the use of embeddings in \techName{} \change{in a supervised way} is a key reason for its highly effective use of transfer learning.

\section{Conclusions}
\label{sec:conclusions}

We have presented \techName{}, a new  powerful approach for matching properties from many sources. It is a machine learning approach that utilizes a large spectrum of features, in particular  embedding features, on both property names and instance values. Our evaluation with four real-world multi-source datasets shows that \techName{} clearly outperforms several baseline approaches  representing the current state-of-the art.  The improvements are even achieved for relatively little training data. Moreover, we showed that the use of embeddings in \techName{} \change{in a supervised way} enables an effective use of transfer learning so that existing high-quality training data from different domains can be utilized to reduce the effort for providing labeled training data. 

In future work, we will investigate the use of \techName{} within a more comprehensive data integration approach for knowledge graphs that also includes entity matching and clustering as well as data fusion. In particular, we plan to evaluate different methods for deriving clusters of equivalent properties from the match results determined with \techName{}.

\section{Acknowledgements}
Our work was supported by the Spanish R\&D\&I programme with grants TIN2016-75394-R, PID2019-105471RB-I00, and P18-RT-1060.

\bibliographystyle{elsarticle-num}
\bibliography{bibliography}
\end{document}